\documentclass[12pt]{article}
\usepackage{amsmath,amsthm, amssymb}
\usepackage{graphicx}
\usepackage{leqno}
\usepackage{hyperref}
\numberwithin{equation}{section}
\begin{document}
\title{\vskip-40pt Response to Wiseman, Rieffel, and Cavalcanti on Bell's 1964 Paper}
\author{Edward J. Gillis\footnote{email: gillise@provide.net}}
				
\maketitle

\begin{abstract} 

\noindent 
Wiseman has claimed that Bell was wrong in stating that determinism was inferred rather than assumed in 
the summary of the EPR argument in his 1964 paper. The reply of Wiseman and his co-authors to my comment 
misstates my reasons for disputing this point, and fails to address the central criticism that their claim 
is based on a seriously flawed formalization of Bell's argument deriving from an unreasonably strong 
interpretation of the the terms, 'influence', 'affect', and 'depend on'.  
\end{abstract}

Wiseman, Cavalcanti, and Rieffel have authored a series of papers\cite{Wiseman_1,Wiseman_2,Wiseman_3} that 
assert that Bell's 1964 paper\cite{Bell_1} assumes, rather than infers, determinism. Their recent 
reply\cite{Wiseman_Reply} to my comment\cite{Gillis_on_WCR} on that assertion in those papers contains several 
misunderstandings of my arguments. In this note I will address only what I regard as the most serious, viz., that 
my disagreement with their claim is based on a failure to clearly distinguish "between  Bell's intuitive views in 
1964 and what he had succeeded in formalizing and proving by that time..." As I stated in my comment "the fundamental 
interpretive error made by WCR" was their seriously flawed formalization of the term, 'influence', used by Bell 
(along with the essentially synonymous terms, 'affect', and 'depend on' in various forms). This flawed formalization 
was the basis for the incorrect definition of 'locality' attributed to Bell, which in turn led to the erroneous 
judgment that Bell's derivation of determinism in the first paragraph of Section 2, was invalid.

I believe that there is general agreement that the brief introductory section of Bell's 1964 paper can be viewed 
as an abstract, and that the body of the argument begins with Section 2. This is consistent with Bell's later (1981) 
characterization of his argument\cite{Bell_Bert}:
\begin{quote}\noindent  
    "My own first paper on this subject[*] starts with a summary of the EPR argument 
    \textit{from locality} to deterministic hidden variables. But the commentators have almost 
    universally reported that it begins with deterministic hidden variables."\end{quote}
Bell appears to be referring here specifically to the opening paragraph of Section 2 of the 1964 paper:
\begin{quotation}\noindent
               ``With the example advocated by Bohm and Aharonov[*], the EPR argument is the following.
                 Consider a pair of spin one-half particles formed somehow in the singlet state and moving 
                 freely in opposite directions. Measurements can be made, say by Stern-Gerlach magnets, on
                 selected components of the spins $\vec{\sigma_1}$ and $\vec{\sigma_2}$. If measurement of 
                 the component $\vec{\sigma_1}\cdot \vec{a}$, where $\vec{a}$ is some unit vector, yields the 
                 value +1 then, according to quantum mechanics, measurement of $\vec{\sigma_2}\cdot \vec{a}$ 
                 must yield the value -1 and vice versa. Now we make the hypothesis[*], and it seems one at 
                 least worth considering, that if the two measurements are made at places remote from one 
                 another the orientation of one magnet does not influence the result obtained with the other. 
                 Since we can predict in advance the result of measuring any chosen component of $\vec{\sigma_2}$, 
                 by previously measuring the same component of $\vec{\sigma_1}$, it follows that the result of 
                 any such measurement must actually be predetermined. Since the initial quantum mechanical wave 
                 function does not determine the result of an individual measurement, this predetermination 
                 implies the possibility of a more complete specification of the state."
              	\end{quotation}
              	
Whether or not one agrees with the validity of the argument in this paragraph, the logical structure is 
clear. Bell describes a situation, makes the assumption that the standard quantum statistical predictions 
apply to possible measurements, formulates a hypothesis, and concludes ("\textit{it follows}") that 
measurement results are predetermined. Wiseman asserts that the argument is invalid\cite{Wiseman_1}: 
\begin{quote}\noindent
"Here Bell has made a mistake. His conclusion (predetermined results) does not follow
from his premises (predictability, and the hypothesis stated in the preceding sentence).
This is simple to see from the following counter-example. Orthodox quantum mechanics
(OQM) is a theory in which the setting a of one device does not statistically influence
the result B obtained with the other: $P_\theta\!(B|a,b,c,\lambda) = P_\theta\!(B|b,c,\lambda).$"\end{quote}

This is the crux of the disagreement. From this point on Bell takes determinism as a consequence of the assumptions 
in his formulation of the EPR argument\cite{EPR}, and uses it as a \textit{premise} in the remainder of the 
paper. His subsequent references to determinism, hidden variables, and similar concepts cannot be taken as evidence 
that his 1964 proof \textit{assumes} determinism. The claim that determinism is a basic assumption of the 
proof, put forth by Wiseman\cite{Wiseman_1}, and endorsed by Cavalcanti\cite{Wiseman_2} and Rieffel\cite{Wiseman_3} 
stands or falls on the question of whether Bell's argument in the opening paragraph of Section 2 is invalid. 

Wiseman's counterexample and claim of invalidity are based on his assertion that Bell has provided a definition 
of 'locality' that can be expressed by the formula quoted above. (As noted by Wiseman this is the formula that 
Jarrett used to define 'locality' two decades later\cite{Jarrett}. This concept was subsequently labeled as 
"parameter independence"(PI) by Shimony\cite{Shimony}, and this is the term most commonly used for it.) In support 
of his assertion he quotes from Bell's Introduction: 
\begin{quote}\noindent
". . . the requirement of locality, or more precisely that the result of a
measurement on one system be unaffected by operations on a distant system
with which it has interacted in the past . . . ."\end{quote}
He also cites a passage from Einstein referred to by Bell. He says that 
\begin{quote}\noindent
"the notions of being "independent of what is done with" [from the Einstein passage] or "unaffected by operations 
on" a system "clearly refer to ...[an] action ...[that] has no statistical effect."\end{quote}
Note that in the previously quoted passage from Wiseman in which he says that Bell has made a mistake, he uses 
PI specifically as a formalization of the phrase, 'does not influence', which is the language used by Bell in the 
paragraph summarizing the EPR argument "...we make the hypothesis...that ...one ... orientation ... does not influence 
the result obtained with the other." Bell uses various forms of the terms, 'affect', 'depend on', and 'influence' more 
or less interchangeably (probably for stylistic reasons), so it seems fair to regard them as equivalent. Therefore, 
we may ask 'Does the formula expressing PI, $P_\theta\!(B|a,b,c,\lambda) = P_\theta\!(B|b,c,\lambda),$" accurately 
capture the meaning of 'does not influence'? A simple example shows that it does not. 

Wiseman applies the terms in the formula as follows. $P$ is the function that assigns probabilities to various 
outcomes, $\theta$ is the theory by which the probabilities are calculated, $B$ represents the outcome of the 
second measurement, $a$ and $b$ represent the settings of the first and second measurement instruments, $c$ represents 
the preparation of the systems to be measured, and $\lambda$ represents all other (possibly hidden) variables that 
might be relevant. But, to assess whether the formula is appropriate for defining 'influence' or similar locutions 
we must view it in more general terms. Abstracting from the particular context, the formula says that the addition of 
the condition, $a$, to the conditions, $(b,c,\lambda)$, does not change the probability of $B$. Wiseman's assertion implies 
that this means that $a$ does not influence $B$. 

Let us apply the formula to a different scenario in which we may talk about influences. Suppose that shortly after 
a very heavy snow storm ($c$) a motorist loses control of his or her vehicle, and gets stuck in a deep snow 
drift off the road. The efforts of the motorist ($b$) to get the vehicle back on the road ($B$) are unsuccessful. 
Another person happens along and attempts to help the stranded motorist ($a$). However, the myriad of surrounding 
conditions ($\lambda$) make it impossible for the two of them to achieve their goal, as could be calculated by 
a thorough, detailed model ($\theta$) taking into account terrain, weather conditions, human capabilities, and other 
relevant factors. In this situation  $P_\theta\!(B|a,b,c,\lambda) = P_\theta\!(B|b,c,\lambda) = 0,$ and so Wiseman 
would say that the help of the passer-by, $a$, has no influence on the desired outcome, $B$. Suppose, now, that 
a second helper, of similar physical abilities to the first comes along and assists ($A$), and together, the three 
people are able to get the vehicle back on the road. This can be represented formally as 
 $P_\theta\!(B|A, a,b,c,\lambda) > P_\theta\!(B|a,b,c,\lambda) = P_\theta\!(B|b,c,\lambda) = 0.$

Despite the successful effort of the trio, according to Wiseman, we are still required to say that the help of the first 
passer-by, $a$, \textit{has no influence} on getting the vehicle back on the road. This is clearly wrong, and this is why 
it is wrong to apply the concept of parameter independence to Bell's argument. Given a set of assumed, fixed, background 
conditions (such as ${b,c,\lambda}$), it is simply not correct to say that for a particular (additional) condition 
(like $a$) to influence an outcome, it must, by itself, change the probability of the outcome. If, in conjunction with 
other additional conditions (including, possibly, outcomes of other actions), it changes the probability then it does have 
an influence. In situations in which a single additional condition, by itself, can change an outcome probability, we would 
typically use a stronger term such as 'determine', or 'cause'. A more reasonable formalization of a weaker 
term like 'influence' (or 'affect' or 'depend on') would be: Given a probability assignment such as $P_\theta\!(B|b,c,\lambda)$), an additional condition, $a$, 
influences outcome, $B$, iff there exists a (possibly empty) set of other conditions, $\{O_1,O_2,...,O_n\}$, such that 
$P_\theta\!(B|O_1,O_2,...,O_n, b,c,\lambda) = P_\theta\!(B|b,c,\lambda)$ and 
$P_\theta\!(B|a,O_1,O_2,...,O_n, b,c,\lambda) \neq P_\theta\!(B|b,c,\lambda)$. This formulation will be used 
below.

If Wiseman were simply claiming that the opening paragraph of Section 2 does not adequately formalize the EPR 
argument to permit an inference of determinism, one might disagree, but it would not be worth debating what is 
essentially a matter of judgment or taste. But Wiseman's claim is much stronger. He argues that it can be 
formalized and, thereby, shown to be invalid. The example just presented shows that his proposed formalization 
is incorrect, and, therefore, his assertion of invalidity cannot be sustained on the basis of it.

The tendency to define 'local' or 'locality' in terms of PI is easy to understand. When the terms in the formula 
are applied appropriately, PI expresses the application of the no-superluminal-signaling principle to the situation 
described by Bell. Since this principle is central to both quantum theory and relativity, and insures compatibility 
between them, it is taken by many as the modern analogue of the principle that waves (or other physical processes) do 
not travel faster than light. Bell discusses this point in "La nouvelle cuisine"\cite{Bell_LNC}, in which he addresses 
various issues surrounding the notion of 'locality'. A great deal of the discussion in this particular debate (on 
all sides\footnote{See also Norsen's analysis\cite{Norsen_on_Wiseman} of Wiseman's first paper.}) has focussed on 
various interpretations of this concept. But a judgment of the validity of Bell's argument in the first paragraph of 
Section 2\cite{Bell_1} must be based on the language he uses there, and it should be noted that he does not use the 
terms 'local', 'locality', 'nonlocal', or any close cognate forms anywhere other than in the Introduction (abstract).

To assess the validity of Bell's argument let us formalize a consequence of his hypothesis that "one ... orientation 
... does not influence the result obtained with the other." Using the modified definition proposed above, one can 
say that it implies that  there does not exist a set of conditions, $\{O_1,O_2,...,O_n\}$, such that \newline 
$P_\theta\!(B|O_1,O_2,...,O_n, b,c,\lambda) = P_\theta\!(B|b,c,\lambda)$ and 
$P_\theta\!(B|a,O_1,O_2,...,O_n, b,c,\lambda) \neq P_\theta\!(B|b,c,\lambda)$. In particular, this entails that  
$P_\theta\!(B|a,A,b,c,\lambda) = P_\theta\!(B|A,b,c,\lambda)$, where $A$ represents the result of the first 
measurement, and the other terms are as described earlier. It should be emphasized that $A$, the result of the 
measurement, is simply $+1$ or $-1$, without any specification of the orientation of the magnet. Since we know that 
$P_\theta\!(B|A,b,c,\lambda) = P_\theta\!(B|b,c,\lambda)$, Bell's hypothesis implies that
$P_\theta\!(B|a,A,b,c,\lambda) = P_\theta\!(B|b,c,\lambda)$. With Bell's stated condition that $ \vec{b} = \vec{a} $, 
the conditions, $a$ and $A$, jointly determine the outcome, $B$. Given his observation that we can choose \textit{any} orientation to measure, it does follow that $ P_\theta\!(B=-A|b,c,\lambda) = 1$ for all $b$. The results are 
predetermined, and Bell's \textit{inference} is valid.

Of course, the expression, $P_\theta\!(B|a,A,b,c,\lambda)$ (with a slight change in the order of the conditions) is 
the one used in Wiseman's statement of Jarrett completeness (aka "Outcome Independence" (OI)): 
$P_\theta\!(B|A,a,b,c,\lambda) = P_\theta\!(B|a,b,c,\lambda)$. As I stated in my earlier comment in describing 
the "fundamental interpretive error" made by the authors, "the negation of either PI or OI would allow a dependence 
of $P(B)$ on the setting, $a$". In other words, the hypothesis that $a$ does not influence $B$ requires both PI and 
OI to be respected. Trivially, this implies the equality derived above: 
$P_\theta\!(B|a,A,b,c,\lambda) = P_\theta\!(B|b,c,\lambda)$. The disagreement concerning the claim that Bell's 
argument is invalid is not a dispute over \textit{whether} it was sufficiently formalized to sustain the conclusion. 
It is about \textit{how} to correctly formalize the language that Bell uses in it. 

The central logical argument just reviewed was embedded in a rather lengthy discussion of what Bell meant by 'locality'. 
That discussion was intended to show that his use of the term 'locality' in both the abstract and in his 1981 
characterization of his EPR summary was appropriate. But, although many might prefer to use the term, 'locality', 
in a more limited sense in contemporary physics,  the critical point is that Bell's derivation of determinism 
from the hypothesis that he explicitly stated in the first paragraph of Section 2 is valid.

\section*{Acknowledgements} 
I would like to thank Travis Norsen for commenting on this paper.

\end{document}